\documentclass[12pt,onecolumn,showpacs,amssymb,aps,nofootinbib,floatfix]{revtex4}
\usepackage{mathtools}
\usepackage{setspace} 
\usepackage{epsfig}
\usepackage{color}
\usepackage{latexsym}
\newcommand{\req}[1]{Eq. \ref{#1}}
\newcommand{\secref}[1]{Section \ref{#1}}
\newcommand{\beq}{\begin{equation}}
\newcommand{\eeq}{\end{equation}}

 \newcommand{\lnz}{\ln \mathcal{Z}}
\newcommand{\bra}[1]{  \left| #1 \right> }
\newcommand{\ket}[1]{  \left< #1 \right| }

\newcommand{\eqcomma}{,\phantom{AA}}
\newcommand{\ave}[1]{\left\langle #1 \right\rangle}
\newcommand{\abs}[1]{\left| #1 \right|}
\begin{document} 
\hbadness=10000
\title{What is ``quantum'' about quantum gravity?}

\author{ Giorgio Torrieri$^a$}
\affiliation{
$^a$ IFGW, Universidade Estadual de Campinas, Campinas, S$\tilde{a}$o Paulo, Brazil\\
{\rm email:torrieri@unicamp.br}\\
%{\rm email addresses: dharam.vir.ahluwalia.1952@gmail.com , lance@phys.ntu.edu.tw , lunogled@gmail.com}\\
%Essay written for the Gravity Research Foundation 2023 Awards for Essays on Gravitation
}

\date{\today}%October 2013}  
\begin{abstract}
  Assuming the validity of the equivalence principle in the quantum regime, we argue that one of the assumptions of the usual definition of quantum mechanics, namely separation between the ``classical'' detector and the ``quantum'' system, must be relaxed.    We argue, therefore, that if both the equivalence principle and quantum mechanics continue to survive experimental tests, that this favors ''informational'' interpretations of quantum mechanics (where formalism is built around relations between observables, defined as information that can be accessed by an observer of the system. In particular ''collapse'' is understood as a change of relative information as the detector interacts with the system ) over ``ontic ones'' (assuming the physical reality of states and wavefunctions, which are assumed to be more than informational objects. In particular collapse is understood as a physical process).   In particular, we show that relational type interpretations can readily accomodate the equivalence principle via a minor modification of the  assumptions used to justify the formalism.
  We qualitatively speculate what a full generally covariant quantum dynamics could look like, and comment on experimental investigations.

%  {\em Essay written for the Gravity Research Foundation 2024 Awards for Essays on Gravitation.}
\end{abstract}
\maketitle
%\newpage
\section{Introduction}
From the first attempts at quantizing gravity \cite{bronstein} it became clear that there must be something fundamentally different about it w.r.t. the other quantum theories.   The basic issue is the equivalence principle\footnote{This term of course can mean several different things.  For the considerations of most of this work the weak non-relativistic form, that the inertial and gravitational mass are equal, is sufficient.  However \secref{relat} defines the equivalence principle as the general covariance of all observables (quantum in this case) which is a stronger form}.  As discussed in \cite{bronstein}, there is tension between quantum canonical uncertainty relations,
\begin{equation}
  \label{uncert}
\ave{\sqrt{\left( \Delta A \right)^2 \left( \Delta B \right)^2} } \geq \frac{1}{2i}\ave{ \abs{ \left[ \hat{A},\hat{B} \right]} }
  \end{equation}
and field theories, because fields can be localized at every point to arbitrary accuracy.   
In natural units for an observable $\hat{P}$ of mass dimension 1 (e.g. the momentum) the full uncertainities will be 
\begin{equation}
\label{uncert2}
\ave{(\Delta P)^2}\sim \underbrace{ \ave{(\Delta P)^2}_{quantum}}_{\sim  \ave{(\Delta x)^{-2}}  \ave{\left[ \hat{P},\hat{x} \right]}}+\frac{Q}{(\Delta x)^2}+\frac{Q}{M (\Delta x)^3}
\end{equation}
where $Q$ is the charge by which the system couples to the apparatus ($\Delta \hat{H}\sim Q q_{detector}$. Also $M$ the system's mass and $\Delta x$ is the system's size including the quantum's uncertainity of its location).
 Physically, the additional terms of \req{uncert2}'s rhs represent additional uncertainities which will not be calculable via an operator algebra of the type of the r.h.s. of \req{uncert}.

However, the first extra term can be brought to be arbitrarily small by considering a ``very large detector'' compared to the system's de Broglie wavelength.   The second extra term is a statement of the system's charge over mass ratio.  In practice this term is the backreaction of the detector to the ``quantum shaking'' of the system and will be small if the system-detector momentum exchange is small w.r.t. the detector's mass.    

In electrodynamics and other standard model theories detectors with arbitrarily low charge to mass ratios can be built so the above issue is irrelevant in comparing quantum calculations to experiment.
Particle detectors are essentially projectors of the momentum operator (the position uncertainity is much larger than the quantum uncertainity) and of course it would be silly to worry about the recoil between the top quark decay products and a CERN experiment.
However, gravity is different: Since inertial mass is equal to the gravitational mass ($\Delta \hat{H} \propto GM m_{system}$ in \req{uncert2}), and since every interaction exchanges momentum and energy, the charge-to-mass ratio of every conceivable detector coupling to any conceivable system is unity, by the equivalence principle.   This statement can be put in a fully relativistic form, since the stress-energy tensor is both the generator of metric transformations (hence 4-momentum current conservation in Minkowski spacetime) and the ``vertex factor'' of graviton emission \cite{megrav,holo}.  Any gravitational interaction can be recast in a non-inertial frame transformation comoving with the detector, with non-inertiality of the order of the graviton momentum. Gravitational systems in superposition,entanglement etc. can likewise be transformed, via a frame redefinition, into detector superpostions \cite{frames}.

Bronstein argued that this implies that, in the domain where gravity is important, quantum mechanics needs a radical modification of the notions of space and time, and perhaps this is the case.  However, the conclusion directly inferred from the calculation is simply that
backreaction of quantum fluctuations  on the ``classical'' detector can never be neglected, and this spoils the canonical quantum results (there is no fully ``classical'' detector and fully ``quantum'' system).   Our understanding of open quantum systems have progressed since \cite{bronstein} was written (see  \cite{caldeira}), so this issue can be examined more quantitatively \cite{megrav}.

One could argue, as does a good part of the modern literature on the topic (the ''holographic approach'', where the gravitational degrees of freedom of a region of spacetime is always encoded on the boundary \cite{harlow}, or the ``S-matrix approach'' where all gravitational observables are assumed to be asymptotic\cite{arkani}) that it is impossible to build a detector capable of probing the quantum theory of gravity.   This might well be true but, as the next \secref{relat} argues, it begs questions which are directly associated with foundations of quantum mechanics rather than quantum gravity.  On the other hand, as \secref{exp} describes, credible proposals not just of idealized detectors but of technologically achievable ones have been made which directly acces observables which are both quantum and gravitational.  Thus, the conudrums described here might be solved in the scientifically best way, via experimental tests. 
\section{Relationalism and unitarity \label{relat}}
\subsection{Bronstein's argument and the quantum formalism \label{relform}}
The main point hidden in Bronstein's argument is that the usual ``mathematical apparatus of quantum mechanics'', describing states as vectors in Hilbert space and observables as operators of this space, essentially assumes that the detector just ``projects out the operator corresponding to the observable'', and in the process ``collapses the wavefunction''.
Many believe these fundamental assumptions of quantum formalism to be ``physical'', and this belief is tied to the belief that quantum states and wavefunctions are ``objective physical objects'' (see for instance \cite{wigner}).
In such a case, Bronstein's argument implies that, provided the equivalence principle is exactly valid, there is new physics beyond quantum mechanics when gravitational effects are present (perhaps of the form \cite{penrose,diosi,curce}), whose mathematical formalization is wholly new, given the ``tight'' axiomatic structure of quantum theory \cite{witalg,witalg2,vicary}.

In fact, putting together recent results on Unruh detectors in a superposition of trajectories \cite{zych,zych2,apadula} (``quantum reference frames'') with the recent surprising argument \cite{wald,wald2} that Rindler horizons collapse quantum superpositions, deviations from quantum mechanics if an ontological interpretation of its formalism is assumed become inevitable even for locally weak gravitational fields, for it would mean superimposed trajectories mean a superposition of timescales at which wavefunction collapse occurs, a paradox if such a collapse is a physical event.

However, others (generally ascribing to QBism, relationalism, the relative state and other ''informational'' formulations) \cite{interp} think that states and wavefunctions are nothing more than convenient mathematical objects to parametrize knowledge accessible to us.  After all, non-commuting operators are inevitable if one assumes that some measurements are inherently incompatible with others; Furthermore, if our theory is to accomodate that
\begin{itemize}
\item Any measurement will yield some answer, represented by  real numbers
\item repeated measurements at vanishing time intervals will yield the same answer 
\end{itemize}
it's inevitable to have, as ingredients of our mathematical formalism 
\begin{description}
\item[Operator eigenvalues] to represent measurement results 
\item[Operator eigenvectors] to represent the system after the result
\item[The operators] must be real-eigenvalued and complete.
\end{description}
Hermitian operators and Hilbert spaces then become the mathematical tool with the required characteristics.

The fundamental principle of this philosophy, as enunciated in \cite{rovelli}, is that what we need for a description of the observable physical world is {\em a physical description of physical systems relative to other systems}, where {\em relative} means some kind of interaction.   The axioms of quantum mechanics clearly fulfills these characterists, but more general axioms could also qualify, as for example open quantum systems described by non-unitary evolutions and with stochastic uncertainities as well as quantum ones.
In this context, then, ''observables'' are properties of the system relevant for the interaction with other systems.  This is again a more general definition than that of usual quantum mechanics, with the latter arising in certain limits (the terms in \req{uncert2} deriving from backreaction are negligible).

In this philosophy, the issues Bronstein raised are "technical" rather than fundamental ones.    One would just have to find more general mathematics objects, with better-suited characteristics in the regime where detector backreaction is significant.
\subsection{Bronstein's argument and relationalism \label{bronrel}}
In particular, a relational philosophy, pioneered in \cite{rovelli}, provides us with a way to ``derive'' what are generally called assumptions of quantum mechanics in a way that can be readily applied to this discussion.
Assuming that  there is a maximum amount of information that may be obtained from a quantum system we may choose a subset $\hat{O}$ of $N$ mutually independent questions, where $N$ is the number of bits contained in the maximum amount of information.  Further assuming that it is always possible to obtain new information from a system, and parametrizing our ignorance in terms of Bayesian probabilities, we could define a probabilistic function $p(\hat{O}_1 |\hat{O})$ parametrizing answers following an observer having obtained the full information accessible of the system.   The Kolmogorov axioms of probability theory force the matrix defined by question $i$ asked via $\hat{O}_1$ to be related to question $j$ asked via $\hat{O}$ to be given in terms of squares of unitary matrices (we use the repeated summation convention in the following expressions)
\begin{equation}
\label{transitivity}
p_{ij} \equiv p(\hat{O}_1^i |\hat{O}^j) = \abs{U_{ij}}^2 \eqcomma U_{ij}U^+_{jk}=\delta_{ik}
\end{equation}
Finally, we must assume transitivity between sets of complete questions.  That is, if $b,c$ are two complete questions, then the unitary matrices associated to them obey $U_{bc}=U_{cd}U_{db}$ for the set of all complete questions $\{d\}$.  This forces the need to represent sets of maximal questions as basis vectors in Hilbert space, and non-compatible sets of maximal questions become linear combinations of such questions. 
\begin{equation}
\label{trans2}
\bra{Q_b}=U_{bc}\bra{Q_c}  \eqcomma p(b|c)=\abs{U_{bc}}^2=\abs{\ket{Q_b}\bra{Q_c}}^2
\end{equation}
  The usual formalism in terms of Hermitian operators and Hilbert spaces follows.

Let us think what happens when the concepts above refer to coordinates, and one expects all quantum observables (i.e. correlators to all orders\footnote{By definition, in a probabilistic theory all observables can be reduced to correlators of a certain order, $C(n)\sim \ave{\phi(x_1)...\phi(x_n)}$. Topological information can be encoded in $\lim_n C(n)$ and in $\lim_{x_i-x_j \rightarrow \infty} C(n)$}) to be generally covariant w.r.t. coordinate changes \cite{frames}.   It will mean that
\begin{equation}
\label{framerel}
 U_{a,b} \rightarrow U_{a,b}^f \eqcomma U^f_{a,b} = F^{+}_{b,c} U_{c,d}^{f'} F_{d,b}
\end{equation}
where the subscript $f$ indicates a frame and  $F$ is going to be an operator associated with a frame transformation from $f$ to $f'$. (not necessarily invertible,as we will see).

It is now worth to compare the above generic definition of relationalism  to the discussion, for example in \cite{witalg}.
The definition of \cite{rovelli} given at the end of \secref{relform} is entirely based on the axioms of probability combined with the limits of information extraction from a system inherent in quantum mechanics.   From this, a set of relations are derived (\req{transitivity} and \req{trans2}) which look exactly like those required for a type I Von Neumann algebra.  \req{framerel} relaxes this assumption by also considering the {\em accessibility} of information to the detector. In \cite{witalg} the argument is made that generically one needs at least a type II algebra  to describe observables in general, since a properly normalized density operator is needed to have well-defined averages and, eventually, compute the entropy.   However a type I algebra (admitting pure states) is only needed for an observer which has full access to a complete time slice, not something that in a locally Lorentz invariant theory happens always.    This means that the interpretation discussion in \cite{rovelli}, when applied to infinite dimensional quantum fields, is fully consistent with the program in \cite{witalg} to use the algebraic properties of quantum fields \cite{witalg2} to construct an algebra of background-independent observations accessible to an observer defined by a worldline.   In the next \secref{bronuni} we examine the physics consequences of this.
\subsection{The equivalence principle instead of unitarity? \label{bronuni}}
At first sight, the formulae in the previous section look like a trivial generalizations of the standard quantum formulae, but it is far from the case.   In \cite{rovelli}, the formalism is build around discreteness (ingrained in $N$ and the finite dimension of $U_{a,b}$).   But coordinate transformations are continuus, necessitating a limit of $N\rightarrow \infty$ while $F$ become a function of the position operator $\hat{x}$ in quantum mechanics (one can define this via a lattice in configuration space).   In quantum field theory, $F$ comes back to be a function of the label $x$ but $U$ becomes an operator valued function of the field operator $\hat{\phi}(x)$. 
As we know, the analytical properties of limits such as $N\rightarrow \infty$ can be different from any finite $N$ properties.

If the frames are non-inertial and/or relativistic several complications arise:
\begin{description}
\item[Causal Horizons] associated with relativistic non-inertial frames.  If the equivalence principle is exact, no contact with the regin beyond the horizon is possible.   Hence,$F_{ij}$ is generally non-unitary and $U^f,U^{f'}$ are in different unitarity classes.  Degrees of freedom beyond the horizon must be traced out by $F_{ij}$ for all questions, a la \cite{laflamme}.   Thus, transitivity as well as  \req{trans2} fails and $F^+ F \ne \delta$.
Note that a time evolution constructed in a metric with a causal horizon should not be a unitary operator, but at most a mixing operator  transforming covariantly under diffeomorphism shifts that exchange causal foliations.  Note that this mixing operator can only be Lindbladian \cite{lind} to leading order, its full non-Markovian structure would be determined by diffeomorphism constraints.  As shown for instance in \cite{diosilind}, a Lindbladian operator is not Lorentz invariant, while \cite{hu} shows it will not have the right zero-temperature (Minkowski) limit.
  \item[Symmetries] vary between inertial and non-inertial frames.  And the concept of symmetries is crucial to the development of quantum theory in terms of representation theory \cite{woit}.   In particular, the Hamiltonian formulation requires a timelike Killing vector field which generally does not exist for non-inertial frames.  
If there is no time-like Killing vector field  \req{transitivity} fails because the sets of all possible questions at different times are not necessarily isomorphic.
Once again this invalidates a unitary representation for $F$s and $U$s (and the non-unitarity of $U$ precludes the existence of a Hermitian Hamiltonian), as well as uncertaity relations of the type reducible to \req{uncert} \cite{way}
  \end{description}
Thus, as described in \cite{witalg} the observables accessible to a detector can not be described by a ``wave-function'' vector in Hilbert space if \req{framerel} holds.  The tension between the definition of local observables, unitarity and diffeomorphism invariance has already been noted in works such as \cite{hao1,hao2}.  If however these observables form a type II algebra, as \cite{witalg} argues for, they can be described by a ``wave-function'' as a parametrization of a non-pure density matrix, 
$ \hat{\rho}_{ij} = \bra{\Psi}_i \prescript{}{j}{\ket{\Psi}}$ (but since there is no Hermitian Hamiltonian there are also no pure states commuting with it) or equivalently a partition function \cite{nishioka} (where general convariance is imposed both on the action and the domain of functional integration, creating a "holographic" constraint \cite{holo}).
The question of what substitutes unitarity as a constraint on operators in the presence of general covariance is of course very much open.   In \cite{megrav} it is speculated that for general covariance unitarity 
$\left( \mathrm{Re}U+\mathrm{Im}U =\left( \mathrm{Re}U-\mathrm{Im}U  \right)^{-1} \right)$ is relaxed to a 
general Kramers-Konig type \cite{tong} fluctuation dissipation relation $\left(
\begin{array}{cc}
\mathrm{Re}\\
\mathrm{Im}
\end{array}U= 
\begin{array}{c}
F_1 ( \mathrm{Im} U)\\
F_2 ( \mathrm{Re} U)
\end{array}
\right)$
for some $F_{1,2}(...)$ determined locally by dynamics, where the "Real" fluctuation term dominates for weak fields (and represents the equivalence between coordinate metric superpositons and detector superpositions) while the "Imaginary" dissipation dominates for stronger fields and represents decoherence from horizons, a la \cite{wald}.
In the operator picture, a dynamically determined fluctuation dissipation relation reflects the well-known fact that the preparation of mixed quantum states is not unique \cite{preskill}.     In this respect the local ``shaking'' of the detector corresponds to incoherent sums of position eigenstates ($\rho \rightarrow \hat{\lambda_i} |x_i(t)><x_i(t)|$ where $\hat{\lambda}$ are classical probabilities of the amount of shaking and  $|x_i(t)><x_i(t)|$ the detector's position in its co-moving frame ) while horizon decoherence represents tracing out of DoFs in the configuration space basis.   In principle, one can see how the two effects could compensate; In other words, if every state is inherently mixed, one could enforce diffeomorphism invariance by demanding that every diffeomorphism represents a state prepatation representing the same density matrix for the causally accessible region by the observer.

This means that Dirac type quantization (behind relations such as \req{uncert}) should not anymore apply:   \req{uncert} is a consequence of Dirac quantization relations \cite{woit}, where canonical pairs of observables have a commutator $i \hbar \delta$ (  The Baker–Campbell–Hausdorff formula implies such a commutator for the Stone-Von Neumann theorem to apply).   Instead, we expect that uncertainity relations to have a component based on commutators (which is related to the uncertainity of another operator), on the detector backreaction (related to the detector's properties) and on the observable's accessible horizon (which is a ''minimum'' bound not related to any other operator, similar to a thermal uncertainity, and in fact interpretable as one via a horizon temperature).

In the partition function approach, the ``violation of canonical quantum mechanics'' can be accomodated as an effective theory, with the charge/mass being the small parameter.  Describing the interaction between the system and the detector via a source term
\begin{equation}
\label{sourceterm}
\mathcal{L}_{int} \simeq  J_i(\tau)x_i \eqcomma \mathcal{L}_{J} \gg \mathcal{L}_{int},\mathcal{L}_{\phi}
\end{equation}
canonical quantum mechanics is equivalent to the part of the partition function originating from $\mathcal{L}_{J}$ is at it's semiclassical value and factorizes from the rest.
\begin{equation}
  \label{zsourceclass}
\lnz \sim \left. \lnz\right|_{J} + \left. \lnz \right|_{system} \eqcomma  \left. \lnz \right|_{system} \ll \left. \lnz\right|_{J}  \eqcomma   \left. \lnz\right|_{J}\simeq \left. \lnz\right|_J^{semi-classical}\simeq \left. S_J \right|_{min}
  \end{equation}
For quantum gravity, provided the equivalence principle holds, this effective theory always breaks down.  However, the semiclassical assumption for the detector can still hold, and in this case, the backreaction can be modeled by a classical-stochastic equation of motion.   Thus, a mixed quantum-classical stochastic theory, of the type recently constructed in \cite{oppenheim} could well emerge not as an extension of quantum mechanics but an effective theory properly incorporating quantum effects and the equivalence principle.   Intriguingly there are indications that the fully relativistic extension of such a theory must be holographic \cite{holo}.
\subsection{A generally covariant ''quantum'' theory: the technical challenge\label{question}}
Thus, the problem of formulating a theory which is both quantum and obeys the equivalence principle exactly has been reformulated as a problem of finding a generally covariant representation for $F,U$ in \req{framerel}.  This is of course a very difficult problem ($F,U$ are infinite dimensional, and $F$ represent coordinate diffeomorphisms).
So far it is fair to say no such theory is conclusively proven to exist.  
The only candidates of quantum gravity explicitly built around the fully quantum equivalence principle, Loop Quantum Gravity and causal dynamical triangulation \cite{lqg,cdt} identify spacetime coordinates (trough area operators,locally quantized via an SU(2) symmetry but transforming covariantly under diffeomorophisms) with observables (so there is no separation between $U$ and $F$) and impose general covariance on the {\em state}, something that is different from the relational discussion in this work, centered around {\em observables} rather than {\em states}\footnote{A direct consequence of diffeomorphism invariance of the quantum state in LQG is the reduction of Unitary time evolution to $\hat{H}|\psi>=0$. We note that the difficoulty of defining time dynamically has given rise to an associated interpretation of quantum measurement \cite{montevideo}, based on assuming any time measurement as involving decoherence \cite{wootters,rovelli}.  The role of general covariance at the quantum level in this interpretation is not clear}.
In realistic physical experiments we do not measure coordinates directly, but infer them via interactions (``rulers'' and ``clocks''), hence the distinction between $U$ and $F$.  General covariance means {\em not} the covariance of a ``state'' (As can be seen, for instance in \cite{kovner} in Non-Abelian Gauge theories wave functionals are explicitly not Gauge-invariant) but of observable correlations (matrices $\abs{U_{ab}}^2$, whose basis are field correlators, under $F-$ type transformations), and general covariance  for field correlators living on a quantum spacetime is not as yet proven for LQG or CDT \cite{lqg,cdt}. What this means for our discussion is that while it is likely that the Algebra inherent in \cite{lqg,cdt} is both of type I and generally covariant, it is not clear what algebra one would get, and if such an algebra would be generally covariant, if one combines the spacetimes of \cite{lqg,cdt} with a field and an observer making observations in such a field.

It is also worth to remark that, to tree level in certain limits quantum field theory already fulfills the requirement of general covariance \cite{matsas} so the hope that this can be brought to higher orders in an EFT expansion, as suggested by \cite{megrav,holo} is not unfounded.  This approach entails reconstructing the metric from QFT observables and seeing what sort of QFT action would lead to a general covariance in terms of this of metric. Such investigations are speculative but ongoing (eg \cite{padma,kempf1,kempf2,tales,nist}).

Most importantly, the problem defining this \secref{question} can be thought of as "technical" rather than "conceptual".  In fact procedures for achieving this have already been proposed \cite{fewster}.   The price for this problem reduction is itself however
 both technical and conceptual in nature.  The technical price is the abandonment of the very convenient Hilbert space and Unitary/Hermitian operator technology.  The conceptual price is a definite abandonment of notions that wavefunctions and states are physical objects.  General covariance makes it inevitable that the only mathematical objects readily associated with a physical meaning are ``sets of questions that can be asked of a system'', i.e. density operators in some observable's basis.     For anyone who does not associate an ontic truth to quantum states, this is a small price to pay, or equivalently a feature rather than a bug.
 \section{An experimental perspective \label{exp}}
 \subsection{Testing the quantum nature of gravity experimentally}
A counter-objection to the discussion here is that perhaps the equivalence principle is broken or even meaningless in a quantum world, as argued in \cite{dono}
Recent proposals for experimental verification of quantum gravity, 
\cite{witness,witness2,gyro,gyro2,pend1} in fact are good laboratories to test
whether the equivalence principle is a good guide to understand post-classical gravity.

\begin{figure}
     \epsfig{width=0.87\textwidth,figure=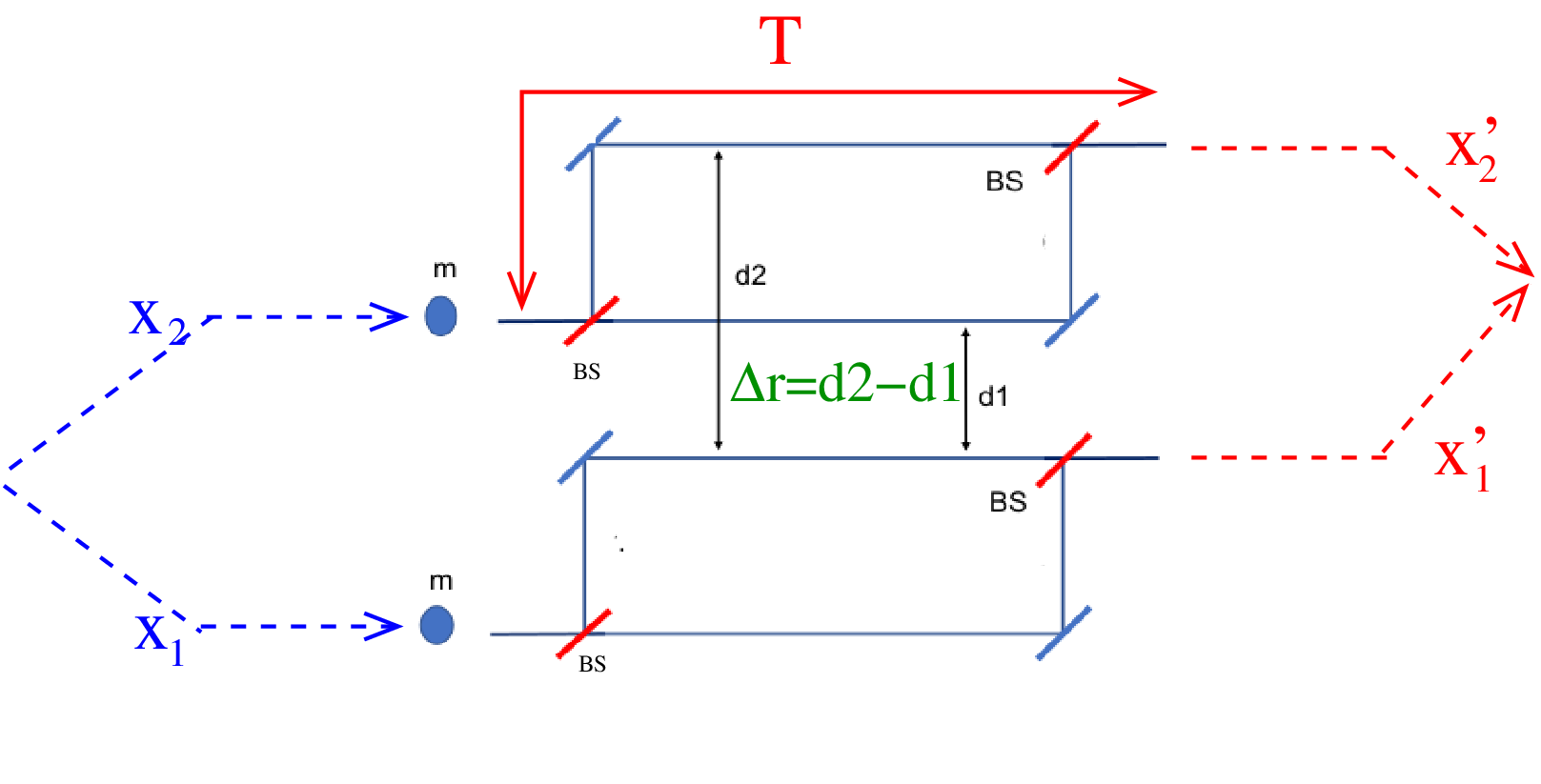}
     \caption{\label{vedral} A double interferometric setup sensitive to the effect of the mutual gravitational interaction of two nanoparticles of mass $m$ on their quantum superposition and entanglement \cite{witness,witness2}.  In the picture we make sure the particles begin and end in the same points, which is not required in \cite{witness,witness2} but becomes important to analyze the covariance properties of the observable, and can be realized by a simple modification of the experiment.}
\end{figure}

For instance in \cite{witness,witness2} (Fig. \ref{vedral})
the authors study the effect on quantum superposition and entanglement of the gravitational interaction between two  nanoparticles of masse $m$, each sent through two superimposed paths in an interferometer using beam splitters (BS).   An observation of the extra phase $\Delta \phi$ in a path difference $T$ of(See Fig. \ref{vedral}) of a "clock" observable tracking time (such as the correlation of entangled precessing spins of the nanoparticles \cite{witness2})
\begin{equation}
  \label{estimate}
\Delta \phi \sim T \Delta E \eqcomma \Delta E \sim \frac{G m^2}{\Delta r}
\end{equation}
would be a proof that the gravitational field is canonically quantized (See also \cite{carney}).

Elsewhere \cite{megrav} we have analyzed this experiment from the point of view of Bronstein's ideas implemented via an effective action of the form \req{sourceterm}, and argued that indeed, while we do not predict a ``negative'' result, the effect of the gravitational decoherence is always comparable to $\Delta \phi$.   If the measuring apparatus can keep two test masses into a superposition of two classical paths, it must mean (just from conservation of momentum) that the apparatus is heavier than the two test masses.
If the mass scale of the detector is $M$ there will be a detector recoil $\sim m/M$ which will dephase \req{estimate}.   
One can of course make $M$ heavier, so $m/M\rightarrow 0$... but then the interaction between $m$ and $M$, $\sim mM$ will be parametrically larger than between the two nanoparticles of mass $m$, which also contributes to dephasing.
\subsection{The Bose-Marletto-Vedral experiment and quantum reference frames}
Looking at the experiment described in the previous section from the point of view of quantum reference frames \cite{frames} indeed confirms that the analysis in \cite{witness,witness2,gyro,gyro2,pend1} is done in the laboratory frame.     Using the spin set-up in \cite{witness}, \req{framerel}'s $U$ will be in the basis of all possible $\ket{\uparrow_1 \uparrow_2(x_{in})}\bra{ \uparrow_1 \downarrow_2(x_{out})}$ where $\uparrow_{1,2}$ transforms as a spinor under the Lorentz group.   Provided the nanoparticles being and end in the same place (something not explicitly required in \cite{witness,witness2} but can be easily accomplished, as Fig. \ref{vedral} shows) the phase change is the difference between two angles propagated through two proper time intervals.   Thus in the classical limit \cite{binney} given a unit vector $A^\mu$  the phase difference $\sim (\Delta A_\mu) A^\mu$ where $\Delta A^\mu= R^\mu_{\alpha \beta \gamma } \Delta S^{\alpha \beta} A^\gamma$ ($A^\mu$ is the vector projection of the difference between two spinors, a vector), where $S^{\alpha \beta}$ is the area element parallel to the interferoemeter and $R^\mu_{\alpha \beta \gamma }$ the Riemann tensor of the gravitational field (sum of earth's, apparatus and nanoparticles).   If tidal forces were to be ignored (they can for the earth, but not the rest) this would of course be a 4-vector.  We are looking, therefore, for the observable to transform covariantly as a 4-vector if general covariance works in the quantum regime.

Now, in the reference frame co-moving with the nano-particles ($f'$) would be a superposition of two $F$ (an up-moving and a down-moving path in Fig. \ref{vedral}), neither of which conserves energy and momentum as the nano-particles recoil.     Thus  $\ket{\uparrow_1 \uparrow_2(x_{in})}\bra{ \uparrow_1 \downarrow_2(x_{out})}$ will not evolve unitarily in $f'$  while of course in the lab frame $f$ the evolution will be unitary.     While an explicit calculation of this is beyond the scope of this work, the unitarity discrepancy means  $\ket{\uparrow_1 \uparrow_2(x_{in})}\bra{ \uparrow_1 \downarrow_2(x_{out})}$ cannot transform as a 4-vector.
Thus the quantum gravity induced phase shift is not generally covariant in the sense of \req{framerel}.    Such covariance would require treating the nano-particles as an open quantum system, interacting with the detector components.   In this case evolution would be non-unitary both in the laboratory and the co-moving frame, but $\Delta A_\mu$ calculated using the partially traced density matrix could in principle transform covariantly.
\section{Conclusion}
The only conclusion we can make with some confidence given the conudrums outlined in section \ref{relat}, and the proposals in section \ref{exp} is that
the study of quantum gravity might be in it's "golden" phase:  Deep conceptual issues in both gravity (the role of the equivalence principle) and quantum mechanics (fundamental limits to the ``classicality'' of the detector and the ``quantumness of the system) might be directly tested by experiments of the type \cite{witness,witness2,gyro,gyro2,pend1}.
If the answer estimated by \cite{witness} will be seen, one can interpret this as a confirmation of the quantization of gravity but also an experimentally verified breakdown of general covariance of quantum observables (or rather of the vacuity of the concept of general covariance in the quantum regime), as advocated in \cite{dono}.
If the result of \cite{witness,witness2,gyro,gyro2,pend1} can not be described by canonical quantum mechanics but rather with some kind of mixed theory a la \cite{oppenheim,megrav,holo} and at the same time stringent tests of the equivalence principle with both astrophysical observations \cite{nordveldt} and tabletop measurements \cite{atoms} yield no violation, one would have to think deeper at how to implement the equivalence principle into a ``relational generally covariant Algebra of observables'' \cite{woit,witalg}, which in general will not be based around unitary matrices. 
The advent of mesoscopic experimental study of quantum effects on gravity, therefore, promises to be important and fruitful.   

\end{document}